\documentclass[titlepage,floatfix,floats,superscriptaddress,repreprint,prl,aps,showpacs,twocolumn]{revtex4-1}
\pdfoutput=1
\usepackage{amssymb}
\usepackage{amsmath}
\usepackage{amsfonts}
\usepackage{graphicx}
\usepackage[mathscr]{euscript}
\usepackage{natbib}

\begin{document}
\title{Conformational Entropy as Collective Variable for Proteins} 
\author{Ferruccio Palazzesi}
\affiliation{Department of Chemistry and Applied Biosciences, ETH Zurich and
            Facolt\`{a} di Informatica, Instituto di Scienze Computationali,
             Universit\`{a} della Svizzera italiana,
             Via Giuseppe Buffi 13, CH-6900, Lugano, Switzerland.}
\author{Omar Valsson}
\affiliation{Department of Chemistry and Applied Biosciences, ETH Zurich and
            Facolt\`{a} di Informatica, Instituto di Scienze Computationali,
             Universit\`{a} della Svizzera italiana,
             Via Giuseppe Buffi 13, CH-6900, Lugano, Switzerland.}
\affiliation{National Center for Computational Design and Discovery of Novel Materials MARVEL.}
             
\author{Michele Parrinello}
\email{parrinello@phys.chem.ethz.ch}
\affiliation{Department of Chemistry and Applied Biosciences, ETH Zurich and
            Facolt\`{a} di Informatica, Instituto di Scienze Computationali,
             Universit\`{a} della Svizzera italiana,
             Via Giuseppe Buffi 13, CH-6900, Lugano, Switzerland.}
\affiliation{National Center for Computational Design and Discovery of Novel Materials MARVEL.}


\date{\today}
\newpage
\begin{abstract}
Many enhanced sampling methods, such as Umbrella Sampling, Metadynamics or Variationally Enhanced Sampling, rely on the identification of appropriate collective variables. For proteins, even small ones, finding appropriate collective variables has proven challenging. Here we suggest that the NMR $S^{2}$ order parameter can be used to this effect.  We trace the validity of this statement to the suggested relation between $S^{2}$ and entropy. Using the $S^{2}$ order parameter and a surrogate for the protein enthalpy in conjunction with Metadynamics or Variationally Enhanced Sampling we are able to reversibly fold and unfold a small protein and draw its free energy at a fraction of the time that is needed in unbiased simulations. From a more conceptual point of view this implies describing folding as a resulting from a trade off between entropy and enthalpy. We also use $S^{2}$ in combination with the free energy flooding method to compute the unfolding rate of this peptide. We repeat this calculation at different temperatures to obtain the unfolding activation energy.
\end{abstract}
\maketitle

Enhanced sampling method have received great attention since they offer the promise of overcoming the limited time scale that direct simulations can afford. Among the plethora of methods proposed in the literature~\cite{voter,Huber1994,helmut,umbrella}, Metadynamics (MetaD)~\cite{Laio,WTMetaD} and more recently Variationally Enhanced Sampling (VES)~\cite{PRL-Valsson2014a} are finding application in a vast array of problems~\cite{palazzesi2013allosteric,martovnak2006crystal,pavan2013combining,branduardi2007b,tiwary2015kinetics,bottaro2016free,shaffer2016enhanced}. Like other similar approaches they rely on the identification of appropriate collective variables (CVs) or order parameters. In MetaD or VES the fluctuations of the CVs are enhanced such that transition between different metastable basins are favored~\cite{BarducciRew,valsson2016enhancing}.

A vast number of CVs have been suggested and are easily accessible in open source codes~\cite{Tribello2014}. However the quest for new order parameters continues in the hope of finding CVs that are efficient and yet generic enough such that they can be applied to a larger class of problems, without prejudging the final results. Identifying appropriate CVs is not only a technical issue needed to accelerate sampling but offers a key to the understanding of the underlying physical processes. The need for such CVs is particularly pressing in the field of bio-molecular systems, such as proteins, whose conformational changes are defined by a large number of degrees of freedom like the arrangement of backbone atoms, side-chains and solvent molecules.

At first this appears like a very demanding request. The purpose of this paper is to show that this is not necessarily so, at least for small proteins or selected regions of larger bio-systems. To this effect we introduce a conceptually new CV and illustrate its efficiency in non trivial examples. Our guiding principle is that in the behavior of proteins, and of many other systems~\cite{piaggi2016enhancing}, entropy plays an important role and if we are able to identify a CV that measures entropy even if in an approximate way, this could go some way towards being able to sample the complex landscape of proteins~\cite{Clementi}. 

In this search we shall be helped by the NMR literature in which several attempts have been made at converting the NMR observable into a measure of conformational entropy~\cite{akke1993nmr,yang1997contributions,stone2001nmr,sharp2015relationship,yang1996contributions,li1996insights,marlow2010role}. Without going into the complex detail of the NMR technique, it suffices to say that in nuclear relaxation experiments, it is possible to measure the dynamical behavior of selected bonds, like N-H or C-H, and, within some approximations, extract the so-called order parameter $S^2$. This parameter, that can vary between 0 and 1, can provide useful information on the degree of spatial motion of the system~\cite{lipari1,lipari2}. 
From the knowledge of this parameter several empirical relation between $S^2$ and the conformational entropy have been proposed and their validity assessed in comparison with either experiments or molecular dynamics simulations~\cite{akke1993nmr,li1996insights,yang1996contributions,sharp2015relationship}. 
Despite being rather empirical~\cite{stone2001nmr}, these relationships have been used to study several bio-macromolecular processes, and to calculate protein heat capacity~\cite{spyracopoulos2001thermodynamic,yang1997contributions,marlow2010role,bracken2001nmr}.

It is the existence of these relations and the notion that entropy plays an important role in proteins that have inspired us to use $S^2$ as CV. However, to order to proceed one needs to express $S^2$ as a function of the atomic coordinates. Luckily such relations are available for both the NH and CH$_3$ groups~\cite{bowman2016accurately,chatfield1998molecular,peter2001calculation,maragakis2008microsecond}. Since here we shall only bias the N-H bonds related order parameter, we solely report the relevant expression proposed by Zhang and Br\"{u}schweiler~\cite{zhang2002contact}:
\begin{equation}
S^{2}=\sum_{n}S^{2}_{n}
\label{EQ:S2}
\end{equation}
where $S^{2}_{n}$ is the order parameter for the $n$-th amino acid residue defined as: 
\begin{equation}
\label{EQ:S2b}
S^{2}_{n}=\tanh\left(0.8\sum_{k}[\exp(-r^{O}_{n-1,k})+\exp(-r^{H_{N}}_{n,k})]\right)-0.1,
\end{equation}
where $k$ runs over all the heavy atoms with the exception of the residues $n$ and $n-1$
and  $r^{H_{N}}_{n,k}$ and  $r^{O}_{n-1,k}$ are the distances from the  heavy atom $k$ from the amide hydrogen in residue $n$ and the carboxyl oxygen in residue $n-1$. In Eq.~(\ref{EQ:S2b}) the distances are to be expressed in {\AA}ngstroms. 

In this paper we assume that this relation is valid for any atomic configuration and we use it as a CV. From a practical point of view the merit of this choice is in the end to be judged by the results. Once the CV is identified, the fluctuations of $S^{2}$ can be amplified by using an enhanced sampling technique, such as MetaD or VES.

\begin{center}
\begin{figure}[h!]
\includegraphics[width=1.0\columnwidth]{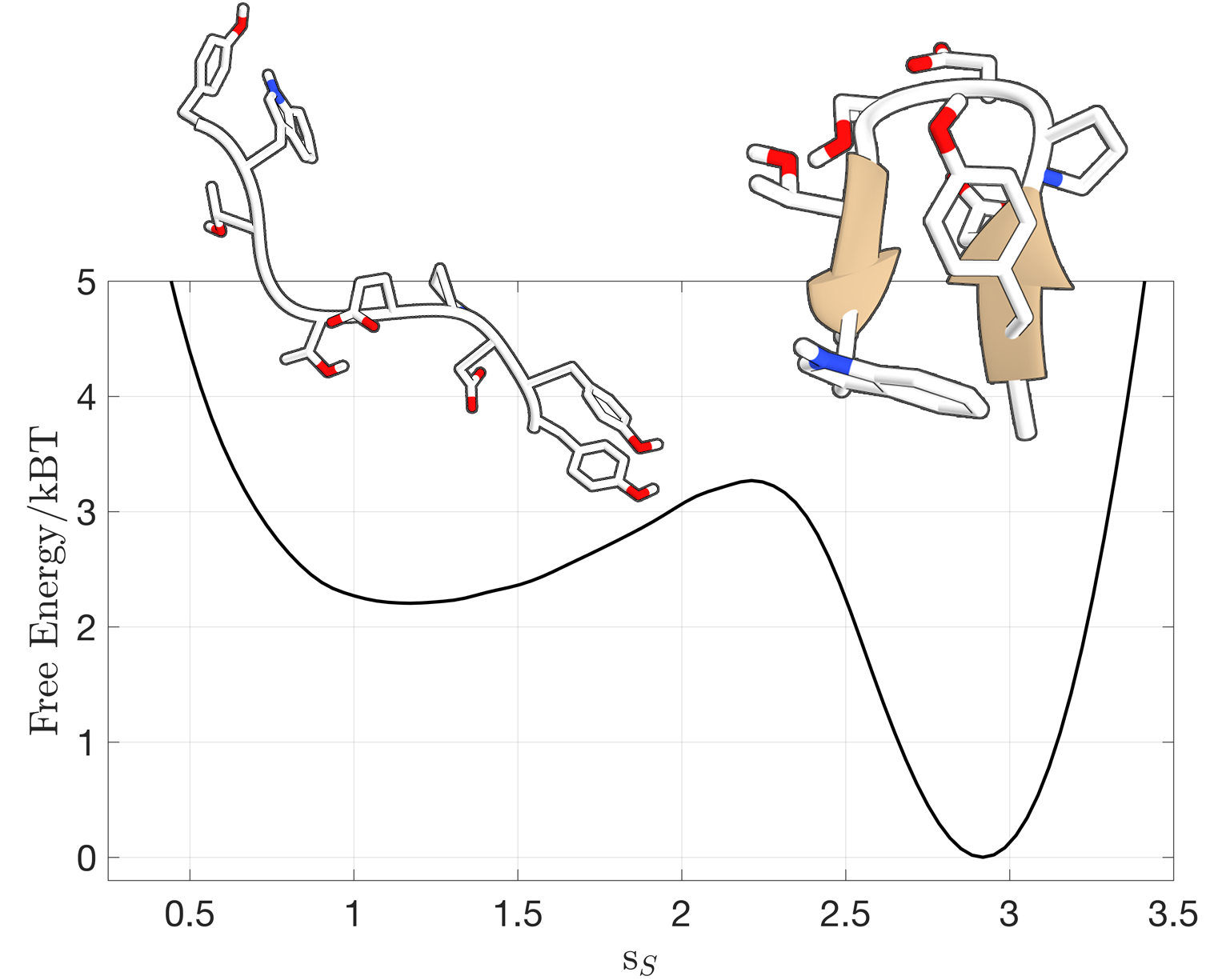}
\caption{FES along the $s_{S}$ CV calculated from long unbiased molecular dynamics simulation of Ref.~\citenum{Science-LindorffLarsen-2011}, with the ribbon representation of chignolin in two representative snapshots of the folded and unfolded conformations.}
\label{FIG:fes}
\end{figure}
\end{center}

Before starting our calculations we indeed checked that $S^{2}$ is able to distinguish between the folded and unfolded protein conformations. For this reason we calculated the free energy surface (FES) along this CV utilizing the chignolin (CLN025)~\cite{honda} trajectory provided to us by the D.E.\ Shaw Research Group~\cite{Science-LindorffLarsen-2011}. As can be seen in Fig.~\ref{FIG:fes}, the FES exhibits two well-defined basins corresponding to the folded and unfolded state. This in itself is an interesting result that suggests the attempt at using $S^{2}$ as collective variable is not totally devoid of merits.

We tried to drive the folding transition using only $S^2$ (hereafter denoted $s_{S}$) as CV but failed. Thus in the spirit of this work in which we describe folding as a trade off between entropy and enthalpy we introduced a second CV that is meant to be a surrogate for enthalpy. This could have been accomplished by separating from the total energy of the systems those components that describe the protein-protein interactions and use these as CV. However this is somewhat expensive and we preferred to use as surrogate for the enthalpy the native H-bonds contact map (hereafter $s_{H}$). At the end we shall reweight the FES~\cite{Rew2} and obtain its projection onto the protein enthalpy defined as the sum of the protein-protein contributions, $E_{pp}$, to the total energy.

We performed the simulation at $T= 340$ K as in Ref.~\citenum{Science-LindorffLarsen-2011} and we use the same potential, the Charmm22*~\cite{c22} protein force field and the TIP3P~\cite{tip3p} water model. We use GROMACS 5.1.4 MD package~\cite{Gromacs5} patched with the PLUMED 2 plug-in~\cite{Tribello2014} in which we have implemented the $s_{S}$ CV. In the MetaD calculations we integrate the equation of motion using a time step of 2.0 fs and the temperature is controlled by the stochastic rescaling thermostat~\cite{Bussi2007a}. Gaussians of initial height 2.82 kJ/mol and width 0.05 for $s_{S}$ and for $s_{H}$ were deposited every picosecond. The value of the bias factor $\gamma$ was set equal to 8~\cite{WTMetaD}. To speed up the calculation and make use of parallelism we used 4 multiple walkers~\cite{Raiteri2006}. We evaluate the convergence using the error metric previously used in Ref.~\cite{WTMetaD,Branduardi2012a,valsson2015well} and using the unbiased data of Ref.~\citenum{Science-LindorffLarsen-2011} as reference.

\begin{center}
\begin{figure}[h!]
\includegraphics[width=1.0\columnwidth]{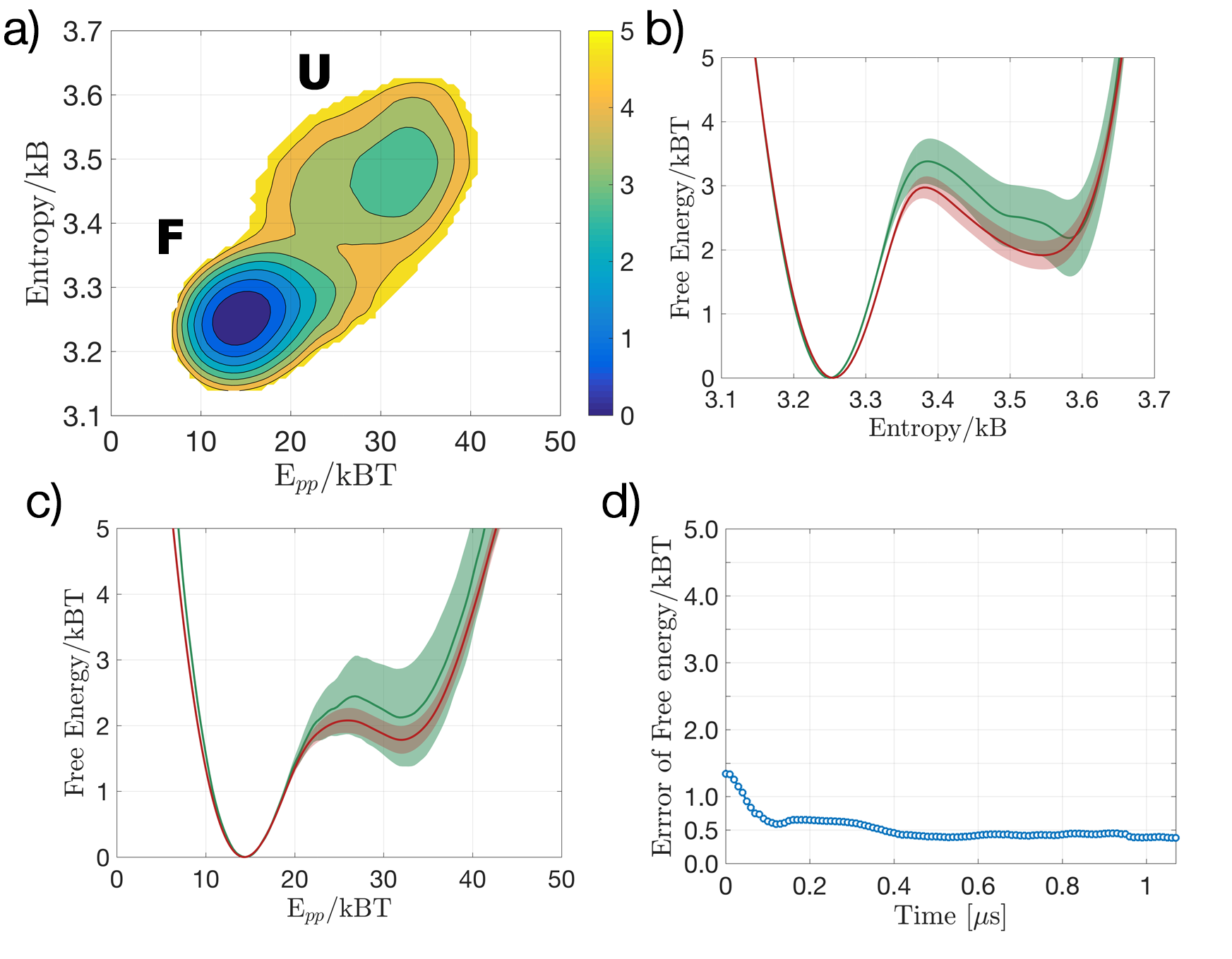}
\caption{a) Reweighted FES calculated from the MetaD simulation using entropy and enthalpy as CVs. b) and c) Mono-dimensional FESs for entropy and enthalpy, respectively. In red the data from the unbiased simulation of Ref.~\citenum{Science-LindorffLarsen-2011}, while in green the one from MetaD simulation. d) Free energy error of the 2D FES along the simulation time. }
\label{FIG:rmsd}
\end{figure}
\end{center}

It can be seen from Fig.~\ref{FIG:rmsd} that the convergence in the FES is reached in about 1.0 $\mu$s, that is a much shorter time relative to $\sim$100.0 $\mu$s reported in Ref.~\citenum{Science-LindorffLarsen-2011}. This reflects the fact that in the MetaD run the rate of transition between folded and unfolded states is accelerated about one hundred times.

In this plot we also express the FES as a function of entropy and enthalpy. The entropy is extimated from s$_{S}$ using the relation given by Wand and co-workers~\cite{sharp2015relationship}, while as a measure of the protein enthalpy we use E$_{pp}$. This calculation clearly proves the usefulness of using entropy and enthalpy to drive reversible transitions between folded and unfolded states.
We have repeated the calculations using VES obtaining statistically indistinguishable results. The VES simulations are performed using the VES code~\cite{vescode} module for PLUMED 2. The results for this calculation are reported in the SM. 

\nocite{Branduardi2012a,valsson2015well}

\begin{center}
\begin{figure}[h!]
\includegraphics[width=1.0\columnwidth]{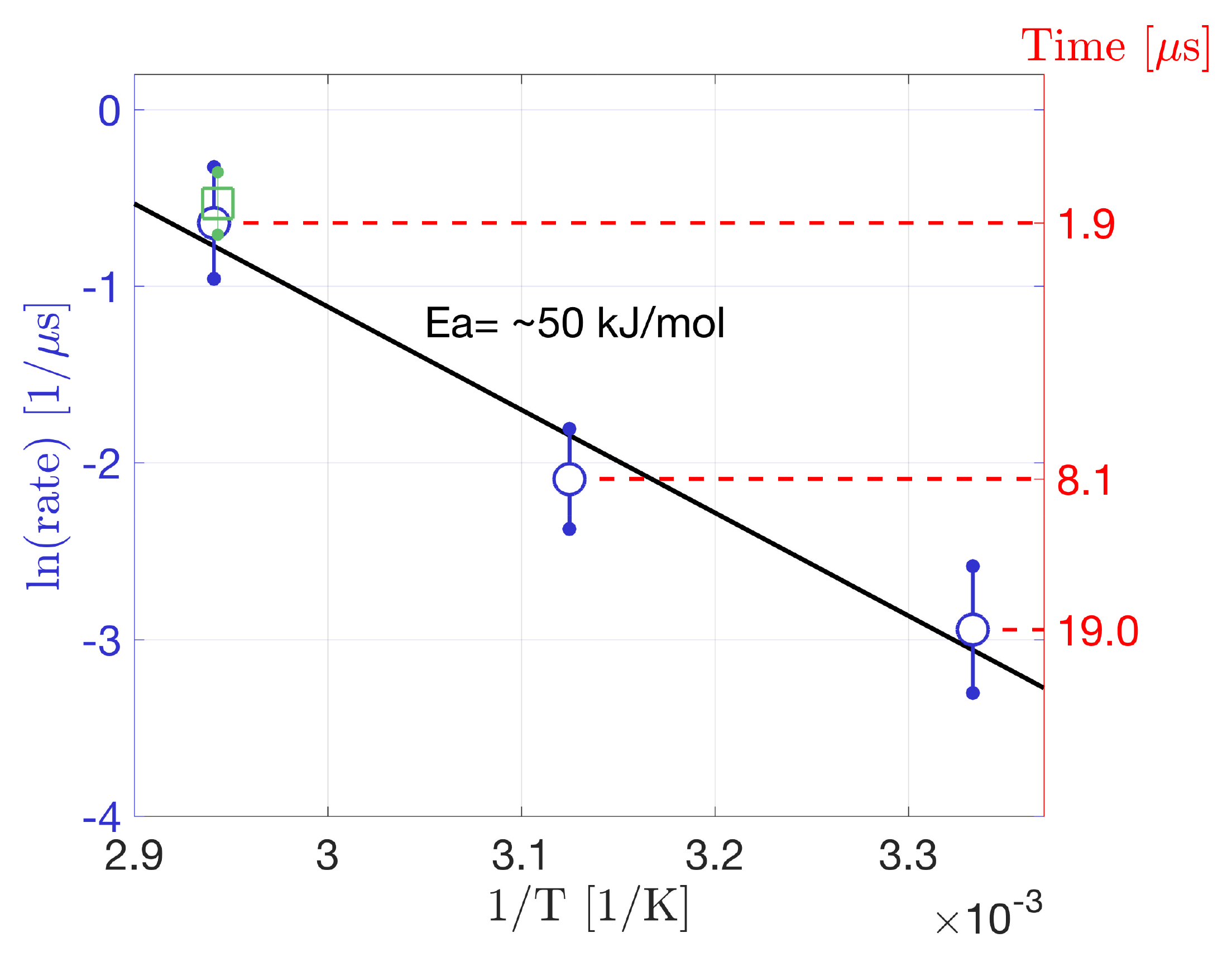}
\caption{Arrenhius plot of the unfolding process of the chignolin. The green square is the unfolding time calculated from the unbiased data of Ref.~\citenum{Science-LindorffLarsen-2011}. The black line is the linear regression used to calculated the activation energy ($R^{2}$ is equal to 0.97). The acceleration factors are 10, 50 and 100 for temperature equal to 340, 320 and 300, respectively. }
\label{FIG:tempi}
\end{figure}
\end{center}

An analysis of the structures lying at the bottom of the folded basin in Fig.~\ref{FIG:fes} shows that they deviate from the experimental structure~\cite{honda} on average by around 0.8 \AA{}. This suggests that if we are only interested in studying the unfolding process, $s_{S}$ alone could be used to promote unfolding transition and calculate unfolding rates using the approach introduced in Ref.~\citenum{mccarty2015variationally}. This is the VES analogous of the infrequent metadynamics method of Tiwary and Parrinello~\cite{dynamics}, that in turn is based on the ideas of Grubmuller~\cite{helmut} and Voter~\cite{voter}. In all these approaches one relates the rates of a rare event, as calculated in a biased system, to the physical unbiased rates by a simple relation. The requirement that the bias does not act in the transition region is crucial for this relation to hold. The methods mentioned earlier differ in the way this is achieved. In Ref.~\citenum{mccarty2015variationally} the variational approach is used to truncate the bias up to a preassigned free energy level. If this cutoff value is smaller than the free energy barrier this latter region remains free of bias and the condition put forward by Grubmuller~\cite{helmut} and Voter~\cite{voter} applies. 
Computational details can be found in the SM. The results of our calculation are shown in Fig.~\ref{FIG:tempi}. The value obtained at $T=340$ K is in good agreement with both the unbiased estimation of Lindorff-Larsen \textit{et al.}~\cite{Science-LindorffLarsen-2011} and the infrequent MetaD simulations performed by Tung and Pfaendtner on a mutated form of our system, using the root-mean square deviation with respect to the folded structure as biased CV~\cite{tung2016kinetics}.

Contrary to these previous estimations we push our calculation to lower temperatures (320 and 300 K), finding for the unfolding of this simple peptide an Arrhenius behavior, with an activation energy of around 50 kJ/mol. This value is in the right ballpark when compared to the estimation based on a similar $\beta$-peptide~\cite{scian2013mutational}.

In conclusion we have shown that the idea of using entropy and enthalpy as collective variables, or rather surrogate expressions for them,  is a useful one. One of the secret of the success in using these CVs is not only that it is founded on physical ideas and concepts but also on the fact that is non-local and thus sensitive to the whole structure of the protein. We would like to add that the $S^{2}$ defined in Eq.~\ref{EQ:S2b} reflects the backbone structure. If one were interested in the role of the side-chains the use of $S^{2}$ order parameter based on the methyl groups might prove useful. 

We believe that here we have introduced a powerful new concept in the simulation of protein.  How far it can be pushed when going to larger systems it remains to be seen. Already at this stage, it can be safely said that, without any modification, the folding of small proteins, the study of intrinsically disordered proteins and the conformational flexibility of proteins segments can be handled as described above~\cite{kokh2016perturbation}. We are confident that with an appropriate adaptation much larger proteins can be similarly handled. 

\nocite{Bach-NIPS-2013,salvalaglio2014assessing}

\begin{acknowledgments}
We acknowledge funding from the European Union Grant No.\ ERC-2014-AdG-670227/VARMET and by the NCCR MARVEL, funded by the Swiss National Science Foundation. This work was supported by a grant from the Swiss National Supercomputing Centre (CSCS) under project ID s721 and u1. The authors thank D.\ E.\ Shaw Research for sharing the simulation data for chignolin.
\end{acknowledgments}

\bibliography{Biblio_PalazzesiValssonParrinello}

\begin{thebibliography}{53}%
\makeatletter
\providecommand \@ifxundefined [1]{%
 \@ifx{#1\undefined}
}%
\providecommand \@ifnum [1]{%
 \ifnum #1\expandafter \@firstoftwo
 \else \expandafter \@secondoftwo
 \fi
}%
\providecommand \@ifx [1]{%
 \ifx #1\expandafter \@firstoftwo
 \else \expandafter \@secondoftwo
 \fi
}%
\providecommand \natexlab [1]{#1}%
\providecommand \enquote  [1]{``#1''}%
\providecommand \bibnamefont  [1]{#1}%
\providecommand \bibfnamefont [1]{#1}%
\providecommand \citenamefont [1]{#1}%
\providecommand \href@noop [0]{\@secondoftwo}%
\providecommand \href [0]{\begingroup \@sanitize@url \@href}%
\providecommand \@href[1]{\@@startlink{#1}\@@href}%
\providecommand \@@href[1]{\endgroup#1\@@endlink}%
\providecommand \@sanitize@url [0]{\catcode `\\12\catcode `\$12\catcode
  `\&12\catcode `\#12\catcode `\^12\catcode `\_12\catcode `\%12\relax}%
\providecommand \@@startlink[1]{}%
\providecommand \@@endlink[0]{}%
\providecommand \url  [0]{\begingroup\@sanitize@url \@url }%
\providecommand \@url [1]{\endgroup\@href {#1}{\urlprefix }}%
\providecommand \urlprefix  [0]{URL }%
\providecommand \Eprint [0]{\href }%
\providecommand \doibase [0]{http://dx.doi.org/}%
\providecommand \selectlanguage [0]{\@gobble}%
\providecommand \bibinfo  [0]{\@secondoftwo}%
\providecommand \bibfield  [0]{\@secondoftwo}%
\providecommand \translation [1]{[#1]}%
\providecommand \BibitemOpen [0]{}%
\providecommand \bibitemStop [0]{}%
\providecommand \bibitemNoStop [0]{.\EOS\space}%
\providecommand \EOS [0]{\spacefactor3000\relax}%
\providecommand \BibitemShut  [1]{\csname bibitem#1\endcsname}%
\let\auto@bib@innerbib\@empty
\bibitem [{\citenamefont {Voter}(1997)}]{voter}%
  \BibitemOpen
  \bibfield  {author} {\bibinfo {author} {\bibfnamefont {A.~F.}\ \bibnamefont
  {Voter}},\ }\href@noop {} {\bibfield  {journal} {\bibinfo  {journal} {Phys.
  Rev. Lett.}\ }\textbf {\bibinfo {volume} {78}},\ \bibinfo {pages} {3908}
  (\bibinfo {year} {1997})}\BibitemShut {NoStop}%
\bibitem [{\citenamefont {Huber}\ \emph {et~al.}(1994)\citenamefont {Huber},
  \citenamefont {Torda},\ and\ \citenamefont {Gunsteren}}]{Huber1994}%
  \BibitemOpen
  \bibfield  {author} {\bibinfo {author} {\bibfnamefont {T.}~\bibnamefont
  {Huber}}, \bibinfo {author} {\bibfnamefont {A.~E.}\ \bibnamefont {Torda}}, \
  and\ \bibinfo {author} {\bibfnamefont {W.~F.}\ \bibnamefont {Gunsteren}},\
  }\href {\doibase 10.1007/bf00124016} {\bibfield  {journal} {\bibinfo
  {journal} {J Computer-Aided Mol Des}\ }\textbf {\bibinfo {volume} {8}},\
  \bibinfo {pages} {695} (\bibinfo {year} {1994})}\BibitemShut {NoStop}%
\bibitem [{\citenamefont {Grubm{\"u}ller}(1995)}]{helmut}%
  \BibitemOpen
  \bibfield  {author} {\bibinfo {author} {\bibfnamefont {H.}~\bibnamefont
  {Grubm{\"u}ller}},\ }\href@noop {} {\bibfield  {journal} {\bibinfo  {journal}
  {Phys. Rev. E}\ }\textbf {\bibinfo {volume} {52}},\ \bibinfo {pages} {2893}
  (\bibinfo {year} {1995})}\BibitemShut {NoStop}%
\bibitem [{\citenamefont {Torrie}\ and\ \citenamefont
  {Valleau}(1977)}]{umbrella}%
  \BibitemOpen
  \bibfield  {author} {\bibinfo {author} {\bibfnamefont {G.~M.}\ \bibnamefont
  {Torrie}}\ and\ \bibinfo {author} {\bibfnamefont {J.~P.}\ \bibnamefont
  {Valleau}},\ }\href@noop {} {\bibfield  {journal} {\bibinfo  {journal} {J.
  Comp. Phys.}\ }\textbf {\bibinfo {volume} {23}},\ \bibinfo {pages} {187}
  (\bibinfo {year} {1977})}\BibitemShut {NoStop}%
\bibitem [{\citenamefont {Laio}\ and\ \citenamefont {Parrinello}(2002)}]{Laio}%
  \BibitemOpen
  \bibfield  {author} {\bibinfo {author} {\bibfnamefont {A.}~\bibnamefont
  {Laio}}\ and\ \bibinfo {author} {\bibfnamefont {M.}~\bibnamefont
  {Parrinello}},\ }\href {\doibase 10.1073/pnas.202427399} {\bibfield
  {journal} {\bibinfo  {journal} {Proc. Natl. Acad. Sci. U.S.A.}\ }\textbf
  {\bibinfo {volume} {99}},\ \bibinfo {pages} {12562} (\bibinfo {year}
  {2002})}\BibitemShut {NoStop}%
\bibitem [{\citenamefont {Barducci}\ \emph {et~al.}(2008)\citenamefont
  {Barducci}, \citenamefont {Bussi},\ and\ \citenamefont
  {Parrinello}}]{WTMetaD}%
  \BibitemOpen
  \bibfield  {author} {\bibinfo {author} {\bibfnamefont {A.}~\bibnamefont
  {Barducci}}, \bibinfo {author} {\bibfnamefont {G.}~\bibnamefont {Bussi}}, \
  and\ \bibinfo {author} {\bibfnamefont {M.}~\bibnamefont {Parrinello}},\
  }\href@noop {} {\bibfield  {journal} {\bibinfo  {journal} {Phys. Rev. Lett.}\
  }\textbf {\bibinfo {volume} {100}},\ \bibinfo {pages} {020603} (\bibinfo
  {year} {2008})}\BibitemShut {NoStop}%
\bibitem [{\citenamefont {Valsson}\ and\ \citenamefont
  {Parrinello}(2014)}]{PRL-Valsson2014a}%
  \BibitemOpen
  \bibfield  {author} {\bibinfo {author} {\bibfnamefont {O.}~\bibnamefont
  {Valsson}}\ and\ \bibinfo {author} {\bibfnamefont {M.}~\bibnamefont
  {Parrinello}},\ }\href {\doibase 10.1103/physrevlett.113.090601} {\bibfield
  {journal} {\bibinfo  {journal} {Phys. Rev. Lett.}\ }\textbf {\bibinfo
  {volume} {113}},\ \bibinfo {pages} {090601} (\bibinfo {year}
  {2014})}\BibitemShut {NoStop}%
\bibitem [{\citenamefont {Palazzesi}\ \emph {et~al.}(2013)\citenamefont
  {Palazzesi}, \citenamefont {Barducci}, \citenamefont {Tollinger},\ and\
  \citenamefont {Parrinello}}]{palazzesi2013allosteric}%
  \BibitemOpen
  \bibfield  {author} {\bibinfo {author} {\bibfnamefont {F.}~\bibnamefont
  {Palazzesi}}, \bibinfo {author} {\bibfnamefont {A.}~\bibnamefont {Barducci}},
  \bibinfo {author} {\bibfnamefont {M.}~\bibnamefont {Tollinger}}, \ and\
  \bibinfo {author} {\bibfnamefont {M.}~\bibnamefont {Parrinello}},\
  }\href@noop {} {\bibfield  {journal} {\bibinfo  {journal} {Proc. Natl. Acad.
  Sci. U.S.A.}\ }\textbf {\bibinfo {volume} {110}},\ \bibinfo {pages} {14237}
  (\bibinfo {year} {2013})}\BibitemShut {NoStop}%
\bibitem [{\citenamefont {Marto{\v{n}}{\'a}k}\ \emph
  {et~al.}(2006)\citenamefont {Marto{\v{n}}{\'a}k}, \citenamefont {Donadio},
  \citenamefont {Oganov},\ and\ \citenamefont
  {Parrinello}}]{martovnak2006crystal}%
  \BibitemOpen
  \bibfield  {author} {\bibinfo {author} {\bibfnamefont {R.}~\bibnamefont
  {Marto{\v{n}}{\'a}k}}, \bibinfo {author} {\bibfnamefont {D.}~\bibnamefont
  {Donadio}}, \bibinfo {author} {\bibfnamefont {A.~R.}\ \bibnamefont {Oganov}},
  \ and\ \bibinfo {author} {\bibfnamefont {M.}~\bibnamefont {Parrinello}},\
  }\href@noop {} {\bibfield  {journal} {\bibinfo  {journal} {Nat. Mat.}\
  }\textbf {\bibinfo {volume} {5}},\ \bibinfo {pages} {623} (\bibinfo {year}
  {2006})}\BibitemShut {NoStop}%
\bibitem [{\citenamefont {Pavan}\ \emph {et~al.}(2013)\citenamefont {Pavan},
  \citenamefont {Barducci}, \citenamefont {Albertazzi},\ and\ \citenamefont
  {Parrinello}}]{pavan2013combining}%
  \BibitemOpen
  \bibfield  {author} {\bibinfo {author} {\bibfnamefont {G.~M.}\ \bibnamefont
  {Pavan}}, \bibinfo {author} {\bibfnamefont {A.}~\bibnamefont {Barducci}},
  \bibinfo {author} {\bibfnamefont {L.}~\bibnamefont {Albertazzi}}, \ and\
  \bibinfo {author} {\bibfnamefont {M.}~\bibnamefont {Parrinello}},\
  }\href@noop {} {\bibfield  {journal} {\bibinfo  {journal} {Soft Mat.}\
  }\textbf {\bibinfo {volume} {9}},\ \bibinfo {pages} {2593} (\bibinfo {year}
  {2013})}\BibitemShut {NoStop}%
\bibitem [{\citenamefont {Branduardi}\ \emph {et~al.}(2007)\citenamefont
  {Branduardi}, \citenamefont {Gervasio},\ and\ \citenamefont
  {Parrinello}}]{branduardi2007b}%
  \BibitemOpen
  \bibfield  {author} {\bibinfo {author} {\bibfnamefont {D.}~\bibnamefont
  {Branduardi}}, \bibinfo {author} {\bibfnamefont {F.~L.}\ \bibnamefont
  {Gervasio}}, \ and\ \bibinfo {author} {\bibfnamefont {M.}~\bibnamefont
  {Parrinello}},\ }\href@noop {} {\bibfield  {journal} {\bibinfo  {journal} {J.
  Chem. Phys.}\ }\textbf {\bibinfo {volume} {126}},\ \bibinfo {pages} {054103}
  (\bibinfo {year} {2007})}\BibitemShut {NoStop}%
\bibitem [{\citenamefont {Tiwary}\ \emph {et~al.}(2015)\citenamefont {Tiwary},
  \citenamefont {Limongelli}, \citenamefont {Salvalaglio},\ and\ \citenamefont
  {Parrinello}}]{tiwary2015kinetics}%
  \BibitemOpen
  \bibfield  {author} {\bibinfo {author} {\bibfnamefont {P.}~\bibnamefont
  {Tiwary}}, \bibinfo {author} {\bibfnamefont {V.}~\bibnamefont {Limongelli}},
  \bibinfo {author} {\bibfnamefont {M.}~\bibnamefont {Salvalaglio}}, \ and\
  \bibinfo {author} {\bibfnamefont {M.}~\bibnamefont {Parrinello}},\
  }\href@noop {} {\bibfield  {journal} {\bibinfo  {journal} {Proc. Natl. Acad.
  Sci. U.S.A.}\ ,\ \bibinfo {pages} {201424461}} (\bibinfo {year}
  {2015})}\BibitemShut {NoStop}%
\bibitem [{\citenamefont {Bottaro}\ \emph {et~al.}(2016)\citenamefont
  {Bottaro}, \citenamefont {Ban{\'a}{\v{s}}}, \citenamefont {Sponer},\ and\
  \citenamefont {Bussi}}]{bottaro2016free}%
  \BibitemOpen
  \bibfield  {author} {\bibinfo {author} {\bibfnamefont {S.}~\bibnamefont
  {Bottaro}}, \bibinfo {author} {\bibfnamefont {P.}~\bibnamefont
  {Ban{\'a}{\v{s}}}}, \bibinfo {author} {\bibfnamefont {J.}~\bibnamefont
  {Sponer}}, \ and\ \bibinfo {author} {\bibfnamefont {G.}~\bibnamefont
  {Bussi}},\ }\href@noop {} {\bibfield  {journal} {\bibinfo  {journal} {J.
  Chem. Phys. Lett.}\ } (\bibinfo {year} {2016})}\BibitemShut {NoStop}%
\bibitem [{\citenamefont {Shaffer}\ \emph {et~al.}(2016)\citenamefont
  {Shaffer}, \citenamefont {Valsson},\ and\ \citenamefont
  {Parrinello}}]{shaffer2016enhanced}%
  \BibitemOpen
  \bibfield  {author} {\bibinfo {author} {\bibfnamefont {P.}~\bibnamefont
  {Shaffer}}, \bibinfo {author} {\bibfnamefont {O.}~\bibnamefont {Valsson}}, \
  and\ \bibinfo {author} {\bibfnamefont {M.}~\bibnamefont {Parrinello}},\
  }\href@noop {} {\bibfield  {journal} {\bibinfo  {journal} {Proc. Natl. Acad.
  Sci. U.S.A.}\ }\textbf {\bibinfo {volume} {113}},\ \bibinfo {pages} {1150}
  (\bibinfo {year} {2016})}\BibitemShut {NoStop}%
\bibitem [{\citenamefont {Barducci}\ \emph {et~al.}(2011)\citenamefont
  {Barducci}, \citenamefont {Bonomi},\ and\ \citenamefont
  {Parrinello}}]{BarducciRew}%
  \BibitemOpen
  \bibfield  {author} {\bibinfo {author} {\bibfnamefont {A.}~\bibnamefont
  {Barducci}}, \bibinfo {author} {\bibfnamefont {M.}~\bibnamefont {Bonomi}}, \
  and\ \bibinfo {author} {\bibfnamefont {M.}~\bibnamefont {Parrinello}},\
  }\href@noop {} {\bibfield  {journal} {\bibinfo  {journal} {WIREs Comput. Mol.
  Sci.}\ }\textbf {\bibinfo {volume} {1}},\ \bibinfo {pages} {826} (\bibinfo
  {year} {2011})}\BibitemShut {NoStop}%
\bibitem [{\citenamefont {Valsson}\ \emph {et~al.}(2016)\citenamefont
  {Valsson}, \citenamefont {Tiwary},\ and\ \citenamefont
  {Parrinello}}]{valsson2016enhancing}%
  \BibitemOpen
  \bibfield  {author} {\bibinfo {author} {\bibfnamefont {O.}~\bibnamefont
  {Valsson}}, \bibinfo {author} {\bibfnamefont {P.}~\bibnamefont {Tiwary}}, \
  and\ \bibinfo {author} {\bibfnamefont {M.}~\bibnamefont {Parrinello}},\
  }\href@noop {} {\bibfield  {journal} {\bibinfo  {journal} {Annual review of
  physical chemistry}\ }\textbf {\bibinfo {volume} {67}},\ \bibinfo {pages}
  {159} (\bibinfo {year} {2016})}\BibitemShut {NoStop}%
\bibitem [{\citenamefont {Tribello}\ \emph {et~al.}(2014)\citenamefont
  {Tribello}, \citenamefont {Bonomi}, \citenamefont {Branduardi}, \citenamefont
  {Camilloni},\ and\ \citenamefont {Bussi}}]{Tribello2014}%
  \BibitemOpen
  \bibfield  {author} {\bibinfo {author} {\bibfnamefont {G.~A.}\ \bibnamefont
  {Tribello}}, \bibinfo {author} {\bibfnamefont {M.}~\bibnamefont {Bonomi}},
  \bibinfo {author} {\bibfnamefont {D.}~\bibnamefont {Branduardi}}, \bibinfo
  {author} {\bibfnamefont {C.}~\bibnamefont {Camilloni}}, \ and\ \bibinfo
  {author} {\bibfnamefont {G.}~\bibnamefont {Bussi}},\ }\href
  {http://dx.doi.org/10.1016/j.cpc.2013.09.018} {\bibfield  {journal} {\bibinfo
   {journal} {Comput. Phys. Commun.}\ }\textbf {\bibinfo {volume} {185}},\
  \bibinfo {pages} {604} (\bibinfo {year} {2014})}\BibitemShut {NoStop}%
\bibitem [{\citenamefont {Piaggi}\ \emph {et~al.}(2016)\citenamefont {Piaggi},
  \citenamefont {Valsson},\ and\ \citenamefont
  {Parrinello}}]{piaggi2016enhancing}%
  \BibitemOpen
  \bibfield  {author} {\bibinfo {author} {\bibfnamefont {P.~M.}\ \bibnamefont
  {Piaggi}}, \bibinfo {author} {\bibfnamefont {O.}~\bibnamefont {Valsson}}, \
  and\ \bibinfo {author} {\bibfnamefont {M.}~\bibnamefont {Parrinello}},\
  }\href@noop {} {\bibfield  {journal} {\bibinfo  {journal} {arXiv preprint
  arXiv:1612.03235}\ } (\bibinfo {year} {2016})}\BibitemShut {NoStop}%
\bibitem [{\citenamefont {Chavez}\ \emph {et~al.}(2004)\citenamefont {Chavez},
  \citenamefont {Onuchic},\ and\ \citenamefont {Clementi}}]{Clementi}%
  \BibitemOpen
  \bibfield  {author} {\bibinfo {author} {\bibfnamefont {L.~L.}\ \bibnamefont
  {Chavez}}, \bibinfo {author} {\bibfnamefont {J.~N.}\ \bibnamefont {Onuchic}},
  \ and\ \bibinfo {author} {\bibfnamefont {C.}~\bibnamefont {Clementi}},\
  }\href {\doibase 10.1021/ja049510+} {\bibfield  {journal} {\bibinfo
  {journal} {J. Am. Chem. Soc.}\ }\textbf {\bibinfo {volume} {126}},\ \bibinfo
  {pages} {8426} (\bibinfo {year} {2004})}\BibitemShut {NoStop}%
\bibitem [{\citenamefont {Akke}\ \emph {et~al.}(1993)\citenamefont {Akke},
  \citenamefont {Brueschweiler},\ and\ \citenamefont
  {Palmer~III}}]{akke1993nmr}%
  \BibitemOpen
  \bibfield  {author} {\bibinfo {author} {\bibfnamefont {M.}~\bibnamefont
  {Akke}}, \bibinfo {author} {\bibfnamefont {R.}~\bibnamefont {Brueschweiler}},
  \ and\ \bibinfo {author} {\bibfnamefont {A.~G.}\ \bibnamefont {Palmer~III}},\
  }\href@noop {} {\bibfield  {journal} {\bibinfo  {journal} {Journal of the
  American Chemical Society}\ }\textbf {\bibinfo {volume} {115}},\ \bibinfo
  {pages} {9832} (\bibinfo {year} {1993})}\BibitemShut {NoStop}%
\bibitem [{\citenamefont {Yang}\ \emph {et~al.}(1997)\citenamefont {Yang},
  \citenamefont {Mok}, \citenamefont {Forman-Kay}, \citenamefont {Farrow},\
  and\ \citenamefont {Kay}}]{yang1997contributions}%
  \BibitemOpen
  \bibfield  {author} {\bibinfo {author} {\bibfnamefont {D.}~\bibnamefont
  {Yang}}, \bibinfo {author} {\bibfnamefont {Y.-K.}\ \bibnamefont {Mok}},
  \bibinfo {author} {\bibfnamefont {J.~D.}\ \bibnamefont {Forman-Kay}},
  \bibinfo {author} {\bibfnamefont {N.~A.}\ \bibnamefont {Farrow}}, \ and\
  \bibinfo {author} {\bibfnamefont {L.~E.}\ \bibnamefont {Kay}},\ }\href@noop
  {} {\bibfield  {journal} {\bibinfo  {journal} {J. Mol. Bio.}\ }\textbf
  {\bibinfo {volume} {272}},\ \bibinfo {pages} {790} (\bibinfo {year}
  {1997})}\BibitemShut {NoStop}%
\bibitem [{\citenamefont {Stone}(2001)}]{stone2001nmr}%
  \BibitemOpen
  \bibfield  {author} {\bibinfo {author} {\bibfnamefont {M.~J.}\ \bibnamefont
  {Stone}},\ }\href@noop {} {\bibfield  {journal} {\bibinfo  {journal} {Acc.
  Chem. Res.}\ }\textbf {\bibinfo {volume} {34}},\ \bibinfo {pages} {379}
  (\bibinfo {year} {2001})}\BibitemShut {NoStop}%
\bibitem [{\citenamefont {Sharp}\ \emph {et~al.}(2015)\citenamefont {Sharp},
  \citenamefont {O'Brien}, \citenamefont {Kasinath},\ and\ \citenamefont
  {Wand}}]{sharp2015relationship}%
  \BibitemOpen
  \bibfield  {author} {\bibinfo {author} {\bibfnamefont {K.~A.}\ \bibnamefont
  {Sharp}}, \bibinfo {author} {\bibfnamefont {E.}~\bibnamefont {O'Brien}},
  \bibinfo {author} {\bibfnamefont {V.}~\bibnamefont {Kasinath}}, \ and\
  \bibinfo {author} {\bibfnamefont {A.~J.}\ \bibnamefont {Wand}},\ }\href@noop
  {} {\bibfield  {journal} {\bibinfo  {journal} {Proteins: Struct., Funct.,
  Bioinf. Proteins}\ }\textbf {\bibinfo {volume} {83}},\ \bibinfo {pages} {922}
  (\bibinfo {year} {2015})}\BibitemShut {NoStop}%
\bibitem [{\citenamefont {Yang}\ and\ \citenamefont
  {Kay}(1996)}]{yang1996contributions}%
  \BibitemOpen
  \bibfield  {author} {\bibinfo {author} {\bibfnamefont {D.}~\bibnamefont
  {Yang}}\ and\ \bibinfo {author} {\bibfnamefont {L.~E.}\ \bibnamefont {Kay}},\
  }\href@noop {} {\bibfield  {journal} {\bibinfo  {journal} {J. Mol. Bio.}\
  }\textbf {\bibinfo {volume} {263}},\ \bibinfo {pages} {369} (\bibinfo {year}
  {1996})}\BibitemShut {NoStop}%
\bibitem [{\citenamefont {Li}\ \emph {et~al.}(1996)\citenamefont {Li},
  \citenamefont {Raychaudhuri},\ and\ \citenamefont {Wand}}]{li1996insights}%
  \BibitemOpen
  \bibfield  {author} {\bibinfo {author} {\bibfnamefont {Z.}~\bibnamefont
  {Li}}, \bibinfo {author} {\bibfnamefont {S.}~\bibnamefont {Raychaudhuri}}, \
  and\ \bibinfo {author} {\bibfnamefont {A.~J.}\ \bibnamefont {Wand}},\
  }\href@noop {} {\bibfield  {journal} {\bibinfo  {journal} {Protein Sci.}\
  }\textbf {\bibinfo {volume} {5}},\ \bibinfo {pages} {2647} (\bibinfo {year}
  {1996})}\BibitemShut {NoStop}%
\bibitem [{\citenamefont {Marlow}\ \emph {et~al.}(2010)\citenamefont {Marlow},
  \citenamefont {Dogan}, \citenamefont {Frederick}, \citenamefont {Valentine},\
  and\ \citenamefont {Wand}}]{marlow2010role}%
  \BibitemOpen
  \bibfield  {author} {\bibinfo {author} {\bibfnamefont {M.~S.}\ \bibnamefont
  {Marlow}}, \bibinfo {author} {\bibfnamefont {J.}~\bibnamefont {Dogan}},
  \bibinfo {author} {\bibfnamefont {K.~K.}\ \bibnamefont {Frederick}}, \bibinfo
  {author} {\bibfnamefont {K.~G.}\ \bibnamefont {Valentine}}, \ and\ \bibinfo
  {author} {\bibfnamefont {A.~J.}\ \bibnamefont {Wand}},\ }\href@noop {}
  {\bibfield  {journal} {\bibinfo  {journal} {Nat. Chem. Bio.}\ }\textbf
  {\bibinfo {volume} {6}},\ \bibinfo {pages} {352} (\bibinfo {year}
  {2010})}\BibitemShut {NoStop}%
\bibitem [{\citenamefont {Lipari}\ and\ \citenamefont
  {Szabo}(1982{\natexlab{a}})}]{lipari1}%
  \BibitemOpen
  \bibfield  {author} {\bibinfo {author} {\bibfnamefont {G.}~\bibnamefont
  {Lipari}}\ and\ \bibinfo {author} {\bibfnamefont {A.}~\bibnamefont {Szabo}},\
  }\href@noop {} {\bibfield  {journal} {\bibinfo  {journal} {J. Am. Chem.
  Soc.}\ }\textbf {\bibinfo {volume} {104}},\ \bibinfo {pages} {4546} (\bibinfo
  {year} {1982}{\natexlab{a}})}\BibitemShut {NoStop}%
\bibitem [{\citenamefont {Lipari}\ and\ \citenamefont
  {Szabo}(1982{\natexlab{b}})}]{lipari2}%
  \BibitemOpen
  \bibfield  {author} {\bibinfo {author} {\bibfnamefont {G.}~\bibnamefont
  {Lipari}}\ and\ \bibinfo {author} {\bibfnamefont {A.}~\bibnamefont {Szabo}},\
  }\href@noop {} {\bibfield  {journal} {\bibinfo  {journal} {J. Am. Chem.
  Soc.}\ }\textbf {\bibinfo {volume} {104}},\ \bibinfo {pages} {4559} (\bibinfo
  {year} {1982}{\natexlab{b}})}\BibitemShut {NoStop}%
\bibitem [{\citenamefont {Spyracopoulos}\ and\ \citenamefont
  {Sykes}(2001)}]{spyracopoulos2001thermodynamic}%
  \BibitemOpen
  \bibfield  {author} {\bibinfo {author} {\bibfnamefont {L.}~\bibnamefont
  {Spyracopoulos}}\ and\ \bibinfo {author} {\bibfnamefont {B.~D.}\ \bibnamefont
  {Sykes}},\ }\href@noop {} {\bibfield  {journal} {\bibinfo  {journal} {Curr.
  Opin. Struct. Bio.}\ }\textbf {\bibinfo {volume} {11}},\ \bibinfo {pages}
  {555} (\bibinfo {year} {2001})}\BibitemShut {NoStop}%
\bibitem [{\citenamefont {Bracken}(2001)}]{bracken2001nmr}%
  \BibitemOpen
  \bibfield  {author} {\bibinfo {author} {\bibfnamefont {C.}~\bibnamefont
  {Bracken}},\ }\href@noop {} {\bibfield  {journal} {\bibinfo  {journal} {J.
  Mol. Graph. Model.}\ }\textbf {\bibinfo {volume} {19}},\ \bibinfo {pages} {3}
  (\bibinfo {year} {2001})}\BibitemShut {NoStop}%
\bibitem [{\citenamefont {Bowman}(2016)}]{bowman2016accurately}%
  \BibitemOpen
  \bibfield  {author} {\bibinfo {author} {\bibfnamefont {G.~R.}\ \bibnamefont
  {Bowman}},\ }\href@noop {} {\bibfield  {journal} {\bibinfo  {journal} {J.
  Comp. Chem.}\ }\textbf {\bibinfo {volume} {37}},\ \bibinfo {pages} {558}
  (\bibinfo {year} {2016})}\BibitemShut {NoStop}%
\bibitem [{\citenamefont {Chatfield}\ \emph {et~al.}(1998)\citenamefont
  {Chatfield}, \citenamefont {Szabo},\ and\ \citenamefont
  {Brooks}}]{chatfield1998molecular}%
  \BibitemOpen
  \bibfield  {author} {\bibinfo {author} {\bibfnamefont {D.~C.}\ \bibnamefont
  {Chatfield}}, \bibinfo {author} {\bibfnamefont {A.}~\bibnamefont {Szabo}}, \
  and\ \bibinfo {author} {\bibfnamefont {B.~R.}\ \bibnamefont {Brooks}},\
  }\href@noop {} {\bibfield  {journal} {\bibinfo  {journal} {J. Am. Chem.
  Soc.}\ }\textbf {\bibinfo {volume} {120}},\ \bibinfo {pages} {5301} (\bibinfo
  {year} {1998})}\BibitemShut {NoStop}%
\bibitem [{\citenamefont {Peter}\ \emph {et~al.}(2001)\citenamefont {Peter},
  \citenamefont {Daura},\ and\ \citenamefont
  {Van~Gunsteren}}]{peter2001calculation}%
  \BibitemOpen
  \bibfield  {author} {\bibinfo {author} {\bibfnamefont {C.}~\bibnamefont
  {Peter}}, \bibinfo {author} {\bibfnamefont {X.}~\bibnamefont {Daura}}, \ and\
  \bibinfo {author} {\bibfnamefont {W.~F.}\ \bibnamefont {Van~Gunsteren}},\
  }\href@noop {} {\bibfield  {journal} {\bibinfo  {journal} {J. Bio. NMR}\
  }\textbf {\bibinfo {volume} {20}},\ \bibinfo {pages} {297} (\bibinfo {year}
  {2001})}\BibitemShut {NoStop}%
\bibitem [{\citenamefont {Maragakis}\ \emph {et~al.}(2008)\citenamefont
  {Maragakis}, \citenamefont {Lindorff-Larsen}, \citenamefont {Eastwood},
  \citenamefont {Dror}, \citenamefont {Klepeis}, \citenamefont {Arkin},
  \citenamefont {Jensen}, \citenamefont {Xu}, \citenamefont {Trbovic},
  \citenamefont {Friesner}, \citenamefont {Palmer~III},\ and\ \citenamefont
  {Shaw}}]{maragakis2008microsecond}%
  \BibitemOpen
  \bibfield  {author} {\bibinfo {author} {\bibfnamefont {P.}~\bibnamefont
  {Maragakis}}, \bibinfo {author} {\bibfnamefont {K.}~\bibnamefont
  {Lindorff-Larsen}}, \bibinfo {author} {\bibfnamefont {M.~P.}\ \bibnamefont
  {Eastwood}}, \bibinfo {author} {\bibfnamefont {R.~O.}\ \bibnamefont {Dror}},
  \bibinfo {author} {\bibfnamefont {J.~L.}\ \bibnamefont {Klepeis}}, \bibinfo
  {author} {\bibfnamefont {I.~T.}\ \bibnamefont {Arkin}}, \bibinfo {author}
  {\bibfnamefont {M.~{\O}.}\ \bibnamefont {Jensen}}, \bibinfo {author}
  {\bibfnamefont {H.}~\bibnamefont {Xu}}, \bibinfo {author} {\bibfnamefont
  {N.}~\bibnamefont {Trbovic}}, \bibinfo {author} {\bibfnamefont {R.~A.}\
  \bibnamefont {Friesner}}, \bibinfo {author} {\bibfnamefont {A.~G.}\
  \bibnamefont {Palmer~III}}, \ and\ \bibinfo {author} {\bibfnamefont {D.~E.}\
  \bibnamefont {Shaw}},\ }\href@noop {} {\bibfield  {journal} {\bibinfo
  {journal} {J. Phys. Chem. B}\ }\textbf {\bibinfo {volume} {112}},\ \bibinfo
  {pages} {6155} (\bibinfo {year} {2008})}\BibitemShut {NoStop}%
\bibitem [{\citenamefont {Zhang}\ and\ \citenamefont
  {Br{\"u}schweiler}(2002)}]{zhang2002contact}%
  \BibitemOpen
  \bibfield  {author} {\bibinfo {author} {\bibfnamefont {F.}~\bibnamefont
  {Zhang}}\ and\ \bibinfo {author} {\bibfnamefont {R.}~\bibnamefont
  {Br{\"u}schweiler}},\ }\href@noop {} {\bibfield  {journal} {\bibinfo
  {journal} {J. Am. Chem. Soc.}\ }\textbf {\bibinfo {volume} {124}},\ \bibinfo
  {pages} {12654} (\bibinfo {year} {2002})}\BibitemShut {NoStop}%
\bibitem [{\citenamefont {Honda}\ \emph {et~al.}(2008)\citenamefont {Honda},
  \citenamefont {Akiba}, \citenamefont {Kato}, \citenamefont {Sawada},
  \citenamefont {Sekijima}, \citenamefont {Ishimura}, \citenamefont {Ooishi},
  \citenamefont {Watanabe}, \citenamefont {Odahara},\ and\ \citenamefont
  {Harata}}]{honda}%
  \BibitemOpen
  \bibfield  {author} {\bibinfo {author} {\bibfnamefont {S.}~\bibnamefont
  {Honda}}, \bibinfo {author} {\bibfnamefont {T.}~\bibnamefont {Akiba}},
  \bibinfo {author} {\bibfnamefont {Y.~S.}\ \bibnamefont {Kato}}, \bibinfo
  {author} {\bibfnamefont {Y.}~\bibnamefont {Sawada}}, \bibinfo {author}
  {\bibfnamefont {M.}~\bibnamefont {Sekijima}}, \bibinfo {author}
  {\bibfnamefont {M.}~\bibnamefont {Ishimura}}, \bibinfo {author}
  {\bibfnamefont {A.}~\bibnamefont {Ooishi}}, \bibinfo {author} {\bibfnamefont
  {H.}~\bibnamefont {Watanabe}}, \bibinfo {author} {\bibfnamefont
  {T.}~\bibnamefont {Odahara}}, \ and\ \bibinfo {author} {\bibfnamefont
  {K.}~\bibnamefont {Harata}},\ }\href@noop {} {\bibfield  {journal} {\bibinfo
  {journal} {J. Am. Chem. Soc.}\ }\textbf {\bibinfo {volume} {130}},\ \bibinfo
  {pages} {15327} (\bibinfo {year} {2008})}\BibitemShut {NoStop}%
\bibitem [{\citenamefont {Lindorff-Larsen}\ \emph {et~al.}(2011)\citenamefont
  {Lindorff-Larsen}, \citenamefont {Piana}, \citenamefont {Dror},\ and\
  \citenamefont {Shaw}}]{Science-LindorffLarsen-2011}%
  \BibitemOpen
  \bibfield  {author} {\bibinfo {author} {\bibfnamefont {K.}~\bibnamefont
  {Lindorff-Larsen}}, \bibinfo {author} {\bibfnamefont {S.}~\bibnamefont
  {Piana}}, \bibinfo {author} {\bibfnamefont {R.~O.}\ \bibnamefont {Dror}}, \
  and\ \bibinfo {author} {\bibfnamefont {D.~E.}\ \bibnamefont {Shaw}},\
  }\href@noop {} {\bibfield  {journal} {\bibinfo  {journal} {Science}\ }\textbf
  {\bibinfo {volume} {334}},\ \bibinfo {pages} {517} (\bibinfo {year}
  {2011})}\BibitemShut {NoStop}%
\bibitem [{\citenamefont {Tiwary}\ and\ \citenamefont
  {Parrinello}(2014)}]{Rew2}%
  \BibitemOpen
  \bibfield  {author} {\bibinfo {author} {\bibfnamefont {P.}~\bibnamefont
  {Tiwary}}\ and\ \bibinfo {author} {\bibfnamefont {M.}~\bibnamefont
  {Parrinello}},\ }\href@noop {} {\bibfield  {journal} {\bibinfo  {journal} {J.
  Phys. Chem. B}\ }\textbf {\bibinfo {volume} {119}},\ \bibinfo {pages} {736}
  (\bibinfo {year} {2014})}\BibitemShut {NoStop}%
\bibitem [{\citenamefont {Piana}\ \emph {et~al.}(2011)\citenamefont {Piana},
  \citenamefont {Lindorff-Larsen},\ and\ \citenamefont {Shaw}}]{c22}%
  \BibitemOpen
  \bibfield  {author} {\bibinfo {author} {\bibfnamefont {S.}~\bibnamefont
  {Piana}}, \bibinfo {author} {\bibfnamefont {K.}~\bibnamefont
  {Lindorff-Larsen}}, \ and\ \bibinfo {author} {\bibfnamefont {D.~E.}\
  \bibnamefont {Shaw}},\ }\href@noop {} {\bibfield  {journal} {\bibinfo
  {journal} {Biophys. J.}\ }\textbf {\bibinfo {volume} {100}},\ \bibinfo
  {pages} {L47} (\bibinfo {year} {2011})}\BibitemShut {NoStop}%
\bibitem [{\citenamefont {Jorgensen}\ \emph {et~al.}(1983)\citenamefont
  {Jorgensen}, \citenamefont {Chandrasekhar}, \citenamefont {Madura},
  \citenamefont {Impey},\ and\ \citenamefont {Klein}}]{tip3p}%
  \BibitemOpen
  \bibfield  {author} {\bibinfo {author} {\bibfnamefont {W.~L.}\ \bibnamefont
  {Jorgensen}}, \bibinfo {author} {\bibfnamefont {J.}~\bibnamefont
  {Chandrasekhar}}, \bibinfo {author} {\bibfnamefont {J.~D.}\ \bibnamefont
  {Madura}}, \bibinfo {author} {\bibfnamefont {R.~W.}\ \bibnamefont {Impey}}, \
  and\ \bibinfo {author} {\bibfnamefont {M.~L.}\ \bibnamefont {Klein}},\
  }\href@noop {} {\bibfield  {journal} {\bibinfo  {journal} {J. Chem. Phys.}\
  }\textbf {\bibinfo {volume} {79}},\ \bibinfo {pages} {926} (\bibinfo {year}
  {1983})}\BibitemShut {NoStop}%
\bibitem [{\citenamefont {Abraham}\ \emph {et~al.}(2015)\citenamefont
  {Abraham}, \citenamefont {Murtola}, \citenamefont {Schulz}, \citenamefont
  {P\'{a}ll}, \citenamefont {Smith}, \citenamefont {Hess},\ and\ \citenamefont
  {Lindahl}}]{Gromacs5}%
  \BibitemOpen
  \bibfield  {author} {\bibinfo {author} {\bibfnamefont {M.~J.}\ \bibnamefont
  {Abraham}}, \bibinfo {author} {\bibfnamefont {T.}~\bibnamefont {Murtola}},
  \bibinfo {author} {\bibfnamefont {R.}~\bibnamefont {Schulz}}, \bibinfo
  {author} {\bibfnamefont {S.}~\bibnamefont {P\'{a}ll}}, \bibinfo {author}
  {\bibfnamefont {J.~C.}\ \bibnamefont {Smith}}, \bibinfo {author}
  {\bibfnamefont {B.}~\bibnamefont {Hess}}, \ and\ \bibinfo {author}
  {\bibfnamefont {E.}~\bibnamefont {Lindahl}},\ }\href {\doibase
  10.1016/j.softx.2015.06.001} {\bibfield  {journal} {\bibinfo  {journal}
  {SoftwareX}\ }\textbf {\bibinfo {volume} {1-2}},\ \bibinfo {pages} {19}
  (\bibinfo {year} {2015})}\BibitemShut {NoStop}%
\bibitem [{\citenamefont {Bussi}\ \emph {et~al.}(2007)\citenamefont {Bussi},
  \citenamefont {Donadio},\ and\ \citenamefont {Parrinello}}]{Bussi2007a}%
  \BibitemOpen
  \bibfield  {author} {\bibinfo {author} {\bibfnamefont {G.}~\bibnamefont
  {Bussi}}, \bibinfo {author} {\bibfnamefont {D.}~\bibnamefont {Donadio}}, \
  and\ \bibinfo {author} {\bibfnamefont {M.}~\bibnamefont {Parrinello}},\
  }\href {\doibase 10.1063/1.2408420} {\bibfield  {journal} {\bibinfo
  {journal} {J. Chem. Phys.}\ }\textbf {\bibinfo {volume} {126}},\ \bibinfo
  {pages} {014101} (\bibinfo {year} {2007})}\BibitemShut {NoStop}%
\bibitem [{\citenamefont {Raiteri}\ \emph {et~al.}(2006)\citenamefont
  {Raiteri}, \citenamefont {Laio}, \citenamefont {Gervasio}, \citenamefont
  {Micheletti},\ and\ \citenamefont {Parrinello}}]{Raiteri2006}%
  \BibitemOpen
  \bibfield  {author} {\bibinfo {author} {\bibfnamefont {P.}~\bibnamefont
  {Raiteri}}, \bibinfo {author} {\bibfnamefont {A.}~\bibnamefont {Laio}},
  \bibinfo {author} {\bibfnamefont {F.~L.}\ \bibnamefont {Gervasio}}, \bibinfo
  {author} {\bibfnamefont {C.}~\bibnamefont {Micheletti}}, \ and\ \bibinfo
  {author} {\bibfnamefont {M.}~\bibnamefont {Parrinello}},\ }\href {\doibase
  10.1021/jp054359r} {\bibfield  {journal} {\bibinfo  {journal} {J. Phys. Chem.
  B}\ }\textbf {\bibinfo {volume} {110}},\ \bibinfo {pages} {3533} (\bibinfo
  {year} {2006})}\BibitemShut {NoStop}%
\bibitem [{\citenamefont {Branduardi}\ \emph {et~al.}(2012)\citenamefont
  {Branduardi}, \citenamefont {Bussi},\ and\ \citenamefont
  {Parrinello}}]{Branduardi2012a}%
  \BibitemOpen
  \bibfield  {author} {\bibinfo {author} {\bibfnamefont {D.}~\bibnamefont
  {Branduardi}}, \bibinfo {author} {\bibfnamefont {G.}~\bibnamefont {Bussi}}, \
  and\ \bibinfo {author} {\bibfnamefont {M.}~\bibnamefont {Parrinello}},\
  }\href@noop {} {\bibfield  {journal} {\bibinfo  {journal} {J. Chem. Theory
  Comput.}\ }\textbf {\bibinfo {volume} {8}},\ \bibinfo {pages} {2247}
  (\bibinfo {year} {2012})}\BibitemShut {NoStop}%
\bibitem [{\citenamefont {Valsson}\ and\ \citenamefont
  {Parrinello}(2015)}]{valsson2015well}%
  \BibitemOpen
  \bibfield  {author} {\bibinfo {author} {\bibfnamefont {O.}~\bibnamefont
  {Valsson}}\ and\ \bibinfo {author} {\bibfnamefont {M.}~\bibnamefont
  {Parrinello}},\ }\href@noop {} {\bibfield  {journal} {\bibinfo  {journal} {J.
  Chem. Theory Comput.}\ }\textbf {\bibinfo {volume} {11}},\ \bibinfo {pages}
  {1996} (\bibinfo {year} {2015})}\BibitemShut {NoStop}%
\bibitem [{ves()}]{vescode}%
  \BibitemOpen
  \href@noop {} {}\bibinfo {note} {\textit{VES Code}, a library that implements
  enhanced sampling methods based on Variationally Enhanced Sampling written by
  O.\ Valsson. For the current version, see
  http://www.ves-code.org}\BibitemShut {NoStop}%
\bibitem [{\citenamefont {McCarty}\ \emph {et~al.}(2015)\citenamefont
  {McCarty}, \citenamefont {Valsson}, \citenamefont {Tiwary},\ and\
  \citenamefont {Parrinello}}]{mccarty2015variationally}%
  \BibitemOpen
  \bibfield  {author} {\bibinfo {author} {\bibfnamefont {J.}~\bibnamefont
  {McCarty}}, \bibinfo {author} {\bibfnamefont {O.}~\bibnamefont {Valsson}},
  \bibinfo {author} {\bibfnamefont {P.}~\bibnamefont {Tiwary}}, \ and\ \bibinfo
  {author} {\bibfnamefont {M.}~\bibnamefont {Parrinello}},\ }\href@noop {}
  {\bibfield  {journal} {\bibinfo  {journal} {Phys. Rev. Lett.}\ }\textbf
  {\bibinfo {volume} {115}},\ \bibinfo {pages} {070601} (\bibinfo {year}
  {2015})}\BibitemShut {NoStop}%
\bibitem [{\citenamefont {Tiwary}\ and\ \citenamefont
  {Parrinello}(2013)}]{dynamics}%
  \BibitemOpen
  \bibfield  {author} {\bibinfo {author} {\bibfnamefont {P.}~\bibnamefont
  {Tiwary}}\ and\ \bibinfo {author} {\bibfnamefont {M.}~\bibnamefont
  {Parrinello}},\ }\href@noop {} {\bibfield  {journal} {\bibinfo  {journal}
  {Phys. Rev. Lett.}\ }\textbf {\bibinfo {volume} {111}},\ \bibinfo {pages}
  {230602} (\bibinfo {year} {2013})}\BibitemShut {NoStop}%
\bibitem [{\citenamefont {Tung}\ and\ \citenamefont
  {Pfaendtner}(2016)}]{tung2016kinetics}%
  \BibitemOpen
  \bibfield  {author} {\bibinfo {author} {\bibfnamefont {H.-J.}\ \bibnamefont
  {Tung}}\ and\ \bibinfo {author} {\bibfnamefont {J.}~\bibnamefont
  {Pfaendtner}},\ }\href@noop {} {\bibfield  {journal} {\bibinfo  {journal}
  {Mol. Sys. Des. Eng.}\ }\textbf {\bibinfo {volume} {1}},\ \bibinfo {pages}
  {382} (\bibinfo {year} {2016})}\BibitemShut {NoStop}%
\bibitem [{\citenamefont {Scian}\ \emph {et~al.}(2013)\citenamefont {Scian},
  \citenamefont {Shu}, \citenamefont {Olsen}, \citenamefont {Hassam},\ and\
  \citenamefont {Andersen}}]{scian2013mutational}%
  \BibitemOpen
  \bibfield  {author} {\bibinfo {author} {\bibfnamefont {M.}~\bibnamefont
  {Scian}}, \bibinfo {author} {\bibfnamefont {I.}~\bibnamefont {Shu}}, \bibinfo
  {author} {\bibfnamefont {K.~A.}\ \bibnamefont {Olsen}}, \bibinfo {author}
  {\bibfnamefont {K.}~\bibnamefont {Hassam}}, \ and\ \bibinfo {author}
  {\bibfnamefont {N.~H.}\ \bibnamefont {Andersen}},\ }\href@noop {} {\bibfield
  {journal} {\bibinfo  {journal} {Biochemistry}\ }\textbf {\bibinfo {volume}
  {52}},\ \bibinfo {pages} {2556} (\bibinfo {year} {2013})}\BibitemShut
  {NoStop}%
\bibitem [{\citenamefont {Kokh}\ \emph {et~al.}(2016)\citenamefont {Kokh},
  \citenamefont {Czodrowski}, \citenamefont {Rippmann},\ and\ \citenamefont
  {Wade}}]{kokh2016perturbation}%
  \BibitemOpen
  \bibfield  {author} {\bibinfo {author} {\bibfnamefont {D.~B.}\ \bibnamefont
  {Kokh}}, \bibinfo {author} {\bibfnamefont {P.}~\bibnamefont {Czodrowski}},
  \bibinfo {author} {\bibfnamefont {F.}~\bibnamefont {Rippmann}}, \ and\
  \bibinfo {author} {\bibfnamefont {R.~C.}\ \bibnamefont {Wade}},\ }\href@noop
  {} {\bibfield  {journal} {\bibinfo  {journal} {J. Chem. Theory Comput.}\
  }\textbf {\bibinfo {volume} {12}},\ \bibinfo {pages} {4100} (\bibinfo {year}
  {2016})}\BibitemShut {NoStop}%
\bibitem [{\citenamefont {Bach}\ and\ \citenamefont
  {Moulines}(2013)}]{Bach-NIPS-2013}%
  \BibitemOpen
  \bibfield  {author} {\bibinfo {author} {\bibfnamefont {F.}~\bibnamefont
  {Bach}}\ and\ \bibinfo {author} {\bibfnamefont {E.}~\bibnamefont
  {Moulines}},\ }in\ \href@noop {} {\emph {\bibinfo {booktitle} {Advances in
  Neural Information Processing Systems 26}}},\ \bibinfo {editor} {edited by\
  \bibinfo {editor} {\bibfnamefont {C.}~\bibnamefont {Burges}}, \bibinfo
  {editor} {\bibfnamefont {L.}~\bibnamefont {Bottou}}, \bibinfo {editor}
  {\bibfnamefont {M.}~\bibnamefont {Welling}}, \bibinfo {editor} {\bibfnamefont
  {Z.}~\bibnamefont {Ghahramani}}, \ and\ \bibinfo {editor} {\bibfnamefont
  {K.}~\bibnamefont {Weinberger}}}\ (\bibinfo  {publisher} {Curran Associates,
  Inc., Red Hook, NY},\ \bibinfo {year} {2013})\ pp.\ \bibinfo {pages}
  {773--781}\BibitemShut {NoStop}%
\bibitem [{\citenamefont {Salvalaglio}\ \emph {et~al.}(2014)\citenamefont
  {Salvalaglio}, \citenamefont {Tiwary},\ and\ \citenamefont
  {Parrinello}}]{salvalaglio2014assessing}%
  \BibitemOpen
  \bibfield  {author} {\bibinfo {author} {\bibfnamefont {M.}~\bibnamefont
  {Salvalaglio}}, \bibinfo {author} {\bibfnamefont {P.}~\bibnamefont {Tiwary}},
  \ and\ \bibinfo {author} {\bibfnamefont {M.}~\bibnamefont {Parrinello}},\
  }\href@noop {} {\bibfield  {journal} {\bibinfo  {journal} {J. Chem. Theory
  Comput.}\ }\textbf {\bibinfo {volume} {10}},\ \bibinfo {pages} {1420}
  (\bibinfo {year} {2014})}\BibitemShut {NoStop}%
\end{thebibliography}%

\end{document}